# Strong enhancement of critical current density in both low & high fields and flux pinning mechanism under hydrostatic pressure in optimally doped (Ba,K)Fe$_2$As$_2$ single crystals

Babar Shabbir[1], Xiaolin Wang[1*], Yanwei Ma[2], Shixue Dou[1], Shishen Yan[3], and Liangmo Mei[3]

[1]Spintronic and Electronic Materials Group, Institute for Superconducting & Electronic Materials, Australian Institute for Innovative Materials, University of Wollongong, Wollongong, NSW, 2500, Australia

[2]Key Laboratory of Applied Superconductivity, Institute of Electrical Engineering, Chinese Academy of Sciences, PO Box 2703, Beijing 100190, China

[3]School of Physics, Shandong University, Shandong, Jinan, P. R. China

Strong pinning depends on the pinning force strength and number density of effective defects. Using hydrostatic pressure method, we demonstrate that hydrostatic pressure up to 1.2 GPa can significantly enhance flux pinning or $J_c$ by a factor of up to 5 especially in both low and high fields in optimally doped Ba$_{0.6}$K$_{0.4}$Fe$_2$As$_2$ crystals. Our analysis on the flux pining mechanism indicate that both pinning centre number density ($N_p$) and pinning force ($F_p$) are greatly increased by the pressure and contribute to strong pinning.

Flux pinning has been a topic of much interest in field of superconductivity because of its importance for applications and fundamental aspects. This interest stems from the significance of flux pinning for high critical current density ($J_c$) in superconductors, which is the defining property of a superconductor. Generally, various types of random imperfections, such as cold-work-induced dislocations [4], secondary-phase precipitates [3], defects induced by high energy ion irradiation [5], etc., can be used to enhance flux pinning. Unfortunately, it is difficult to discern the maximum potential of a superconductor from these techniques, and the outcomes hold up only to a certain level. Furthermore, the critical current is only enhanced, in most of cases, either in low or high fields, with degradation of the superconducting critical temperature ($T_c$) another drawback. For instance, proton irradiation can only enhance flux pinning in high fields by inducing point defects in Ba122:K. Similarly, light ion C$^{4+}$ irradiation of Ba122:Ni crystals can only enhance $J_c$ in low fields at high temperatures [1]. High energy particle irradiation can also decrease $T_c$ by more than 5 K for cobalt and nickel doped Ba-122 [2, 3].

As is well known, $J_c$ is mostly limited by weak links (in the case of polycrystalline bulks), and thermally activated flux creep (an intrinsic property) emerges from weak pinning [4-10]. Strong pinning can be achieved either by inducing effective pinning centres with strong pinning force.

Our previous results show that $J_c$ is enhanced significantly under hydrostatic pressure at high fields (i.e., over one order of magnitude) in comparison to low fields, along with enhancement of the closely related $T_c$ by more than 5 K in Sr$_4$V$_2$O$_6$Fe$_2$As$_2$ polycrystalline bulks and NaFe$_{0.97}$Co$_{0.03}$As single crystals [11, 12]. Until now, however, it has been unclear that the observed $J_c$ enhancement under pressure is correlated with improved $T_c$ or flux pinning. The primary motivation for the present work is to use optimally doped single crystal samples (which has an unchanged $T_c$ under hydrostatic pressure), to elucidate the contributions of flux pinning to $J_c$ enhancement in Fe-based superconductors. The secondary motivation is to investigate further the contributions from both $N_p$ and $F_p$ to strong pinning.

The argument is as follows: Hydrostatic pressure can induce pinning centres, which, in turn, enhance the pinning force. The total pinning force and the pinning centres are correlated by $F_p = N_p f_p$ where $N_p$ is the number density of pinning centres and $f_p$ is the elementary pinning force, which is the maximum pinning strength of an individual pinning centre, with a value that depends on the interaction of the flux line with the defect. According to the flux pinning theory, strongly interacting defects can contribute to $F_p$ individually, provided that $F_p \propto N_p$, and weakly interacting defects can contribute only collectively; the collective theory leads to $F_p \propto (N_p)^2$ for small defect numbers [13].

The Ba122:K compound is believed to be the most technologically suitable because of its isotropic nature and high $T_c$, upper critical field ($H_{c2}$), and $J_c$ values ($J_c > 10^6$ A/cm$^2$ at 2 K and 0 T) [14-18]. The depairing current density ($J_d$) is the maximum current density that superconducting electrons can support before de-pairing of Cooper pairs, and is given as

$$J_d = \frac{\Phi_o}{3\sqrt{3}\pi\mu_o\lambda^2\xi} \quad (1)$$

The $J_d$ value that is found is roughly 0.3 GA/cm$^2$ by using values of penetration depth, $\lambda = 105$ nm and coherence length, $\xi = 2.7$ nm [19, 20]. Our estimation indicates that there is a significant potential to further enhance flux pinning in the (Ba,K)Fe$_2$As$_2$.

In this paper, we investigate the flux pinning of optimally doped (Ba,K)Fe$_2$As$_2$ under hydrostatic pressure. We demonstrate that hydrostatic pressure causes little change in $T_c$, but leads to significant enhancement in flux pinning or $J_c$ by a factor of 10 in both low and high fields in optimally doped Ba$_{0.6}$K$_{0.4}$Fe$_2$As$_2$ crystals. Our analysis shows that the both $N_p$ and $F_p$ are increased by the pressure and contribute to strong pinning.

High quality 122 crystals were grown by using a flux method. The pure elements Ba, K, Fe, As, and Sn were mixed in a mol ratio of Ba$_{1-x}$K$_x$Fe$_2$As$_2$:Sn = 1:45–50 for the self-flux. A crucible with a lid was used to reduce the evaporation loss of K as well as that of As during growth. The crucible was sealed in a quartz ampoule filled with Ar and loaded into a box furnace. The temperature dependence of the magnetic moments and the *M-H* loops at different temperatures and pressures were measured on a Quantum Design Physical Property Measurement System (QD PPMS 14 T) by using the Vibrating Sample Magnetometer (VSM). We used an HMD high pressure cell and Daphne 7373 oil as a pressure transmitting medium to apply hydrostatic pressure on the samples.

Figure 1 shows the temperature dependence of the magnetic moments for zero-field-cooled (ZFC) and field-cooled (FC) measurements at different pressures. $T_c$ remains almost unchanged at different pressures. $T_c \approx$ 37.95 K was found at $P = 0$ GPa and $P = 1$ GPa. Similar results were also reported for $Ba_{0.6}K_{0.4}Fe_2As_2$ thin film [21]. Furthermore, a temperature independent magnetic moment at low temperatures was observed, along-with a small transition width, indicating the high quality of the crystals.

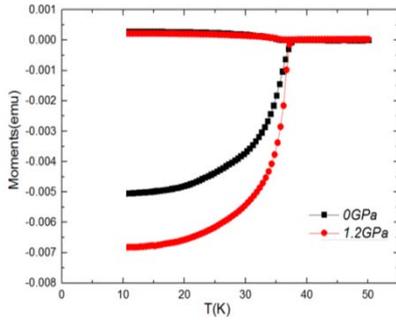

Figure 1: Magnetic moments versus temperature at $P = 0$ GPa and $P = 1.2$ GPa

The field dependence of $J_c$ at different temperatures (4.1, 16, and 24 K) and pressures (0 and 1.2 GPa), obtained from the *M-H* curves by using Bean's model, are shown in Fig. 2. Nearly five-fold $J_c$ enhancement can be seen at 16 K and 24 K in both low and high fields at P=1.2GPa. It is noteworthy that $J_c$ is enhanced for the $Ba_{0.6}K_{0.4}Fe_2As_2$ crystal at 1.2 GPa in both low and high fields. This has not been found by other approaches for pinning enhancement reported so far. At 16 K and self-field, the $J_c$ is $2 \times 10^5$ and increases up to $6 \times 10^5$ A/cm$^2$ by pressure of 1.2GPa and retains as high as $3 \times 10^5$ A/cm$^2$ at 12 T. At 24 K, $J_c$ at zero field is $9 \times 10^4$ A/cm$^2$ and raises up to $2.5 \times 10^5$ A/cm$^2$ at P=1.2GPa and remains the same level at 12 T. At 4.1 K, the $J_c$ is nearly $1 \times 10^6$ A/cm$^2$ at 2 T and $5 \times 10^5$ A/cm$^2$ at 12 T under P=1.2GPa.

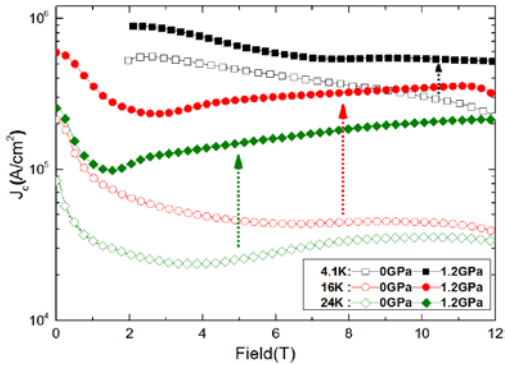

Figure 2: $J_c$ as a function of field at $P = 0$ and 1.2 GPa at 4.1, 16, and 24 K.

The pinning force ($F_p = J_c \times B$) as a function of field at 8 K, 12 K, 24 K, and 28 K is plotted in Figure 3 [22]. At high fields and pressures, the $F_p$ is found to be nearly 5 times higher at 8, 12, 24, and 28 K as compared to $P = 0$ GPa, which corresponds nicely to $J_c$ enhancement. Figure 4 shows a comparison of $F_p$ obtained in our $Ba_{0.6}K_{0.4}Fe_2As_2$ under pressure with those of several other low and high temperature superconducting materials [23-26]. The (Ba,K)Fe$_2$As$_2$ shows better in-field performance under pressure. Pressure can significantly improve $F_p$ values to greater than 60 GN/m$^3$ at $H > 10$ T, which are even superior to those of Nb$_3$Sn and NbTi.

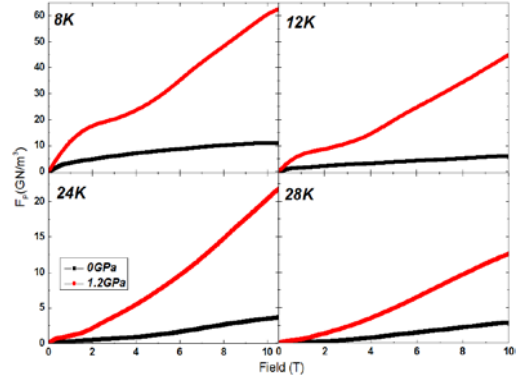

Figure 3: $F_p$ versus field at 8, 12, 24, and 28 K at different pressures.

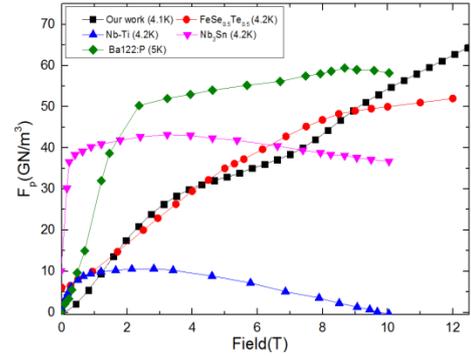

Figure 4: Comparison of $F_p$ for different superconductors.

With respect to the $N_p$, pressure can also increase the number of point pinning centres, which can suppress thermally activated flux creep, leading to $J_c$ enhancement [11]. The $N_p$ can be calculated from the following relation:

$$\frac{\Sigma F_p}{\eta f_p^{max}} = N_P \quad (2)$$

Where $\Sigma F_p$ is the accumulated pinning force density, $f_p^{max}$ is the maximum elementary pinning force $(f_p)$, which is the interaction between a flux line and a single defect, and $\eta$ is an efficiency factor. $\eta = 1$ corresponds to a plastic lattice, and the $\eta$ value is otherwise $f_p^{max}/B$, where $B$ is the bulk modulus of the material. We assume to a second order of approximation that the interaction between a flux line and a single defect is nearly same under pressure. Therefore, we can use $f_p^{max} \approx 3 \times 10^{-13}$ N for a similar superconductor (i.e., Ba122:Co) to estimate $N_p$ [27]. At 4.1 K, $N_p \approx 7.3 \times 10^{24}$ m$^{-3}$ at $P = 0$ Gpa, which increased to $N_p \approx 1.2 \times 10^{25}$ m$^{-3}$ for $P = 1.2$ GPa, while at 24 K, $N_p \approx 6.6 \times 10^{23}$ m$^{-3}$ at $P = 0$ Gpa, which increased to $N_p \approx 3.8 \times 10^{24}$/m$^3$ for $P = 1.2$ GPa.

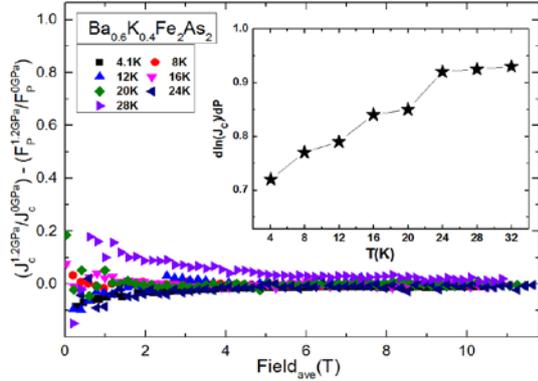

Figure 5: $J_c$-$F_p$ ratios at $P = 1.2$ GPa and $P = 0$ GPa. The relative change of $J_c$ with pressure as a function of $T$ is given in the inset.

In order to examine if the pinning force enhancement is a major factor responsible for the observed $J_c$ enhancement in our crystal under pressure, we have calculated the difference in ratio of $J_c^{1.2GPa}/J_c^{0GPa}$ ($J_c^r$) and $F_P^{1.2GPa}/F_P^{0GPa}$ ($F_p^r$) and plot result in Fig. 5 as a function of field. Analysis of the $J_c^r - F_p^r$ data, acquired at different temperatures, leads to values of nearly zero. This result indicates that $J_c$ enhancement is only related to pinning force enhancement.

To examine whether the observed $J_c$ enhancement is likely affected by volume change of the samples under high pressure, we have performed the following analysis. According to the Wentzel-Kramers-Brillouin (WKB) approximation, high pressure can modify grain boundaries through reduction of the tunnelling barrier width and the tunnelling barrier height for polycrystalline bulks, correlated to following simple mathematical expression [28-30]:

$$J_c = J_{co}\exp(-2kW) \quad (3)$$

Where $W$ is the barrier width, $k = (2mL)^{1/2}/\hbar$ is the decay constant, which is barrier height ($L$) dependent, $\hbar$ is the reduced Planck constant, and $J_{c0}$ is the critical current density at 0 K and 0 T. The relative pressure dependence of $J_c$ can be determined from Eq. (1) as [31]:

$$\frac{d\ln J_c}{dP} = \frac{d\ln J_{c0}}{dP} - \left[\left(\frac{d\ln W}{dP}\right)\ln\left(\frac{J_{c0}}{J_c}\right)\right] - \frac{1}{2}\left[\left(\frac{d\ln L}{dP}\right)\ln\left(\frac{J_{c0}}{J_c}\right)\right]$$

$$= \frac{d\ln J_{c0}}{dP} + \kappa_{GB}\ln\left(\frac{J_{c0}}{J_c}\right) + \frac{1}{2}\kappa_L\ln\left(\frac{J_{c0}}{J_c}\right) \quad (2)$$

Where the compressibility in the width and height of the grain boundary are defined by $\kappa_{GB} = -d\ln W/dP$ and $\kappa_L = -d\ln L/dP$, respectively.

For the (Ba,K)Fe$_2$As$_2$ single crystals, we assume to a first approximation that $\kappa_{GB}$ and $\kappa_L$ are nearly comparable to the average linear compressibility values $\kappa_a = -d\ln a/dP$ ($\kappa_a \approx 0.00318$ GPa$^{-1}$) and $\kappa_c = -d\ln c/dP$ ($\kappa_c \approx 0.00622$ GPa$^{-1}$), respectively, in the FeAs plane, where $a$ and $c$ are the in-plane and out-of-plane lattice parameters [32]. Therefore, we can write Eq. (3) correspondingly as

$$\frac{d\ln J_c}{dP} \approx \frac{d\ln J_{c0}}{dP} + \kappa_a \ln\left(\frac{J_{c0}}{J_c}\right) + \frac{1}{2}\kappa_c \ln\left(\frac{J_{c0}}{J_c}\right) \quad (4)$$

By using $J_c \approx 10^5$ A/cm$^2$ at 24 K and $J_{c0} \approx 10^6$ A/cm$^2$, we find that ($\kappa_a \ln(J_{co}/J_c)$) $\approx 0.0073$ GPa$^{-1}$ and ($1/2\ \kappa_c \ln(J_{co}/J_c)$) $\approx 0.0071$ GPa$^{-1}$, so both only contributed less than 2% to the experimental value, i.e., $d\ln J_c/dP = 0.92$ GPa$^{-1}$ from the inset of Figure 5. This illustrates that the origin of the strong flux pinning under pressure does not arise from the change in volume.

The $J_c$ value as a function of reduced temperature (i.e. 1-$T/T_c$) at 0 and 10 T under different pressures is shown in Fig. 6. The data points in different fields and pressures follow a power law description [i.e. $J_c \propto (1-T/T_c)^\beta$], where $\beta$ is a critical exponent [33-35]. Ginzburg-Landau theory predicts different vortex pinning mechanisms at specified fields, with different values of exponent $\beta$. It was found that $\beta = 1$ corresponds to non-interacting vortices and $\beta \geq 1.5$ refers to the core pinning mechanism. The exponent $\beta$ (i.e., slope of the fitting line) is found to be 1.74 and 1.85 for zero field, and 1.20 and 1.43 at 10 T, at 0 and 1.2 GPa, respectively, which reveals a strong $J_c$ dependence on pressure. The low values of $\beta$ at high pressure show that the $J_c$ decays rather slowly in comparison to its values at low pressure. Different values of exponent $\beta$ have also been observed in MgB$_2$ and yttrium barium copper oxide (YBCO) [36, 37]

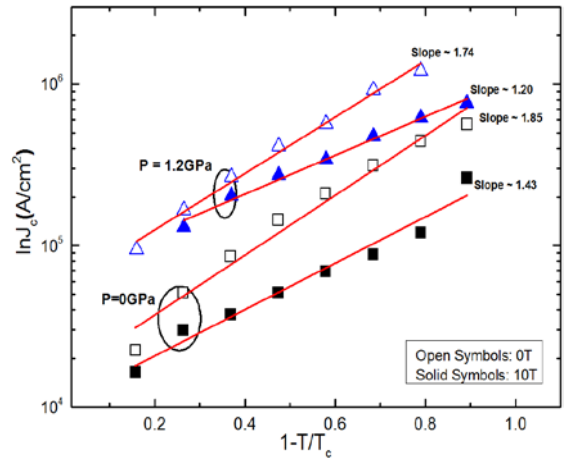

Figure 6: ln$J_c$ versus reduced temperature at different fields and pressures.

The pinning mechanisms in $Ba_{0.6}K_{0.4}Fe_2As_2$ have been analysed by using collective pinning theory. Generally, core pinning comprises 1) $\delta\ell$ pinning, which comes from spatial variation in the charge carrier mean free path, $\ell$, and 2) $\delta T_c$ pinning due to randomly distributed spatial variation in $T_c$.

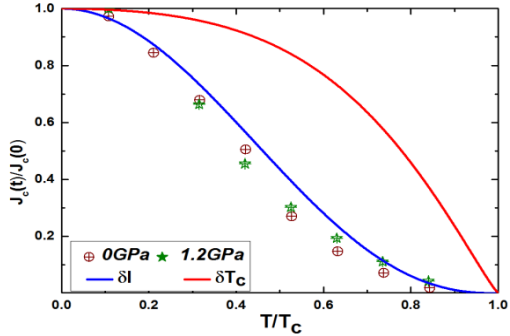

Figure 7: $J_c$ as a function of $T/T_c$. Experimental data points are in agreement with **$\delta\ell$** pinning.

Referring to the theoretical approach proposed by Griessen et al.:

$$J_c(t)/J_c(o) \propto (1-t^2)^{5/2}(1+t^2)^{-1/2} \quad (5)$$

in the case of $\delta\ell$ pinning, while

$$J_c(t)/J_c(0) \propto (1-t^2)^{7/6}(1+t^2)^{5/6} \quad (6)$$

applies in the case of $\delta T_c$ pinning, where $t = T/T_c$ [38]. Fig.7 shows almost perfect overlapping of the experimentally obtained $J_c$ values and the theoretically expected variation for the $\delta\ell$ pinning mechanism at 0.05 T. This is in agreement with the observation of little change in $T_c$ under high pressure. We also observed similar results in $BaFe_{1.9}Ni_{0.1}As_2$ and $SiCl_4$ doped $MgB_2$ [39, 40]. Furthermore, $\delta\ell$ pinning has also been reported in $FeTe_{0.7}Se_{0.3}$ crystals [41].

We have studied the flux pinning in optimally doped $Ba_{0.6}K_{0.4}Fe_2As_2$ crystal under hydrostatic pressure, analysing the critical current density determined experimentally. We propose that strong flux pinning in a wide range of fields can be achieved by improving the pinning force under pressure. The pressure of 1.2 GPa improved the $F_p$ by nearly 5 times at 8, 12, 24, and 28 K, which can increase $J_c$ by nearly two-fold at 4.1 K and five-fold at 16 K and 24 K in both low and high fields, respectively. This study also demonstrates that such an optimally doped superconductor's performance in both low and high fields can also be further enhanced by pressure.